# Two-Axis planar Hall magnetic field sensors with sub nanoTesla resolution

P.T. Das[1,2,#], H. Nhalil[1,#,*], V.Mor[1], M.Schultz[1], N. Hasidim[3], A. Grosz[3], and L. Klein[1,*]

[1]Department of Physics, Institute of Nanotechnology and Advanced Materials, Bar-Ilan University, Ramat-Gan 529002, Israel
[2]Intelligent Materials and Systems (FWID), Helmholtz Center Dresden-Rossendorf, Bautzner Landstraße 400 – 01328 Dresden, Germany
[3]Department of Electrical and Computer Engineering, Ben-Gurion University of the Negev, P.O. Box 653, Beer-Sheva 84105, Israel

**Planar Hall effect (PHE) magnetic sensors are attractive for various applications where the field resolution is required in the range of sub-nano Tesla or in Pico Tesla. Here we present a detailed noise study of the PHE sensors consisting of two or three intersecting ellipses. It can be used to measure two axes of the magnetic field in the sensor plane in particular along the two perpendicular easy axes in the overlapping region for two intersecting ellipses and three easy axes at an angle of 60 degrees for three crossing ellipses. Thus, for each remanent magnetic state in the overlap area, the sensor can measure the vector component of the magnetic field perpendicular to the direction of the remanent magnetization. The two field components are measured with a field resolution ≤ 200 pT/√Hz at 10 Hz and 350 pT/√Hz at 1 Hz in the same region, while maintaining a similar size and noise level of a single-axis sensor. Furthermore, we discuss here the possible route for future improvement of the field resolution.**

## I. INTRODUCTION

Magnetic sensors play an essential role in a large number of scientific and industrial areas including medicine [1-3], space exploration, military [4], [5], archaeology [6], [7], etc. Among the wide variety of room temperature magnetic sensors available based on magnetoresistance (MR) effects, planar Hall effect (PHE) sensors are particularly attractive for low noise measurements as their equivalent magnetic noise is comparable and even surpass the best commercial MR magnetometers [8-10]. Furthermore, compared to sensors based on the AMR effect, giant magnetoresistance (GMR) effect or tunnelling magnetoresistance (TMR) effect, PHE sensors have intrinsically linear low field response, small hysteresis and high thermal stability making them suitable for the substitution of many of the MR sensors. Additionally, they are relatively easy to manufacture and theretofore potentially cheaper.

The PHE is closely related to the AMR effect and it yields a transverse resistivity $\rho_{xy}$ which is given by,

$$\rho_{xy} = \frac{1}{2}\Delta\rho \, sin(2\varphi) \qquad (1)$$

where, $\Delta\rho = \rho_{||} - \rho_{\perp}$, $\rho_{||}$ and $\rho_{\perp}$ are the resistivities parallel and perpendicular to the magnetization, respectively. $\varphi$ is the angle between the bias current and the magnetization.

Our previous studies have demonstrated that elliptical PHE sensors behave much like a single magnetic domain due to shape induced anisotropy along the long axis [11]. It has also been established that two and three crossing ellipses can have four and six stable magnetic states respectively [12]. Here, we report noise studies of Permalloy based PHE sensors with two or three crossing ellipses that measure two axes of the in-plane field with a sensor resolution or equivalent magnetic noise (EMN) level below 200 pT/√Hz at 10 Hz and 350 pT/√Hz at 1 Hz.

## II. EXPERIMENTAL DETAILS

Permalloy ($Ni_{80}Fe_{20}$) films capped with tantalum (Ta) on Si substrates are deposited in an Ultra-high vacuum (UHV) evaporation and sputtering system [13]. The two and three crossing elliptical sensors are patterned using photolithography in a lift-off process. The size of each ellipse is 5 mm in length and 0.650 mm in width with a thickness of 200 nm. Gold leads and contact pads are deposited in the second stage. The noise spectral density (NSD) of the sensors is measured in a bandwidth of 0.01-20 Hz. We excite the sensors with an ac current generated by a function generator (PXI-5421, National Instruments). The sensor signal is modulated to a high frequency by a NI-PXIe-4464 function generator and amplified using a transformed matched amplifier (TMA). Details of the TMA can be found elsewhere [11]. The TMA output is sampled (PXI-4464, National Instruments) and demodulated using a digital synchronous detector. The output signal is filtered to a bandwidth of 100 Hz by low pass filter of the synchronous detector. Since the TMA has a white input spectral noise above 1 kHz, the excitation frequency is kept above this frequency (1.22 kHz) to avoid the TMA's 1/f noise and the power-line harmonics (50 Hz). All measurements are performed in a tri-layer magnetic shield (made of Amumetal) at room temperature. Note that during the measurements the sensor is excited with an alternating current of optimized amplitude and frequency [15] and before the noise measurement a magnetic field of 10 mT is temporarily (for a few seconds) applied along the easy-axis (EA) of interest to initialize a uniform remanent magnetization along the EA.

[#]*Equal contributions*; *Corresponding Authors

## III. RESULTS AND DISCUSSION

Fig.1 exhibits schematics of two and three crossing ellipses sensors along with the measurement geometry. For two crossing ellipses, there are two easy axes, EA1 and EA2, perpendicular to each other as shown in Fig.1(a) and making an angle 45° with the major axes of the ellipses which are the hard axes for this geometry. This structure has four stable magnetic states as reported earlier [14]. The magnetic behavior in the overlap region of the ellipses is described by the effective Hamiltonian,

$$\mathcal{H} = K_2 \cos^2 2\theta - M_s H \cos(\beta - \theta) \quad (2)$$

where, $K_2$ is the biaxial anisotropy constant, $\beta$ is the angle between the magnetic field $H$ and the major axis, $\theta$ is the angle between the magnetization and the major axis, and $M_s$ is the saturation magnetization.

In case of three crossing ellipses, there are 3 easy axes [12], along the major axes of the ellipses that are separated by 60° (see EA1, EA2 and EA3 in Fig.1(b)). This structure has six stable remanent magnetic states and the magnetic behavior in the overlap area is described by effective Hamiltonian [12],

$$\mathcal{H} = K_3 \sin^2 3\theta - M_s H \cos(\beta - \theta) \quad (3)$$

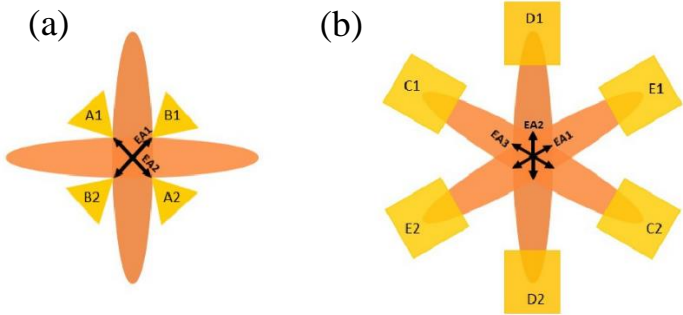

Fig. 1. Typical patterns of (a); two crossing ellipses and (b); three crossing ellipses. The black two headed arrows (EA1, EA2 and EA3) mark the easy axes and the yellow-colored regions are the gold contact pads. For sensors with crossing ellipses, the excitation is driven between a pair of leads along the direction of the remanent magnetization (A1-A2 or B1-B2) and the PHE signal is measured between the other pair. For sensors with 3-ellipses, the current is driven between a pair of leads along the direction of the remanent magnetization (C1-C2, D1-D2 or E1-E2) and the PHE signal is measured between the leads across the current. For example, if the excitation current is between D1-D2, the PHE signal can be measured either across C1-E1 or between E2-C2.

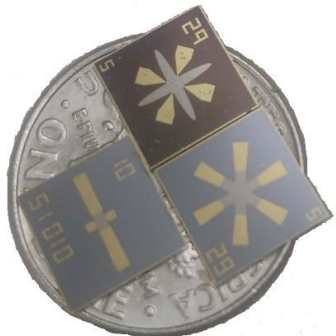

Fig. 2. Photograph of one, two and three crossing ellipses sensors on ONE DIME as size reference.

where, $K_3$ is the triaxial anisotropy constant, $\beta$ is the angle between the magnetic field $H$ and the major axis, $\theta$ is the angle between the magnetization and the major axis, and $M_s$ is the saturation magnetization.

The sensitivity ($S_y$) of a PHE sensor is the ratio of the PHE voltage ($V_y$) and the magnetic field $B$ applied in the film plane perpendicular to the easy axis [15]. For a magnetic field much smaller than the magnetic anisotropy, $S_y$ is given by,

$$S_y = \frac{V_y}{B} = 10^4 \frac{V_x}{R_x} \cdot \frac{\Delta\rho}{t} \cdot \frac{1}{H_k + H_a} \quad (4)$$

here, $V_x$ and $R_x$ are the bias voltage and resistance between the x-terminals (along easy axis of the sensor), $\Delta\rho$ is the average sensor resistivity ($\Delta\rho = \rho_\parallel - \rho_\perp$), $t$ is the film thickness, $H_k$ and $H_a$ are the sensor shape induced and intrinsic anisotropy, respectively. In case of two or three crossing ellipse sensors, the excitation direction is taken as the x-direction and the direction perpendicular to it is taken as the y-direction (hard axis of the sensor).

The sensor resolution is determined by its equivalent input noise and by its sensitivity (Eq. 4). The total NSD of a PHE sensor is composed of contributions from two major noise components: 1/f noise and thermal noise (white noise). In addition to these two intrinsic noise components, an extrinsic noise contribution from TMA (preamplifier noise) also contributes to the NSD. The, total NSD is thus given by,

$$e_\Sigma = \sqrt{V_x^2 \frac{\delta_H}{N_C \cdot vol \cdot f^\alpha} + 4k_B T R_y + e_{amp}^2} \quad (5)$$

where, $\delta_H$ is the Hooge parameter [16], $N_C$ is the free electron density of Permalloy (1.7 x 10$^{29}$ /m$^3$), Vol is the effective volume of the sensor where the electrons contribute to the conduction process in a homogeneous sample, $f$ is the frequency of the excitation, $\alpha$ is a constant on the order of 1, $k_B$ is the Boltzmann constant, $T$ is the temperature, $R_y$ is the sensor resistance perpendicular to the excitation-line, and $e_{amp}$ is the total preamplifier noise of the TMA [16]. Thus, the sensor equivalent magnetic noise between two voltage sensing pads, $B_{eq} (= \frac{e_\Sigma}{S_y})$ is defined as a ratio between total estimated noise of the system and estimated signal level of the sensor (sensitivity).

In case of two crossing ellipses, when the magnetization is along one of the easy axes (EA1 or EA2, Fig.1(a)), we measured the component of the magnetic field perpendicular to the remanent magnetization. Thus, if the remanent magnetization is along EA1(EA2), by exciting the sensor along A1-A2 (B1-B2) the external field component perpendicular to EA1 (EA2) can be measured by measuring the PHE signal across B1-B2 (A1-A2) perpendicular to the remanent magnetization. Here, Fig. 2 exhibits the photograph of real sensors on ONE DIME as a size reference.

Fig. 3 presents the $B_{eq}$ vs frequency for excitation along A1-A2 and measurement across B1-B2 (see Fig. 3(a)) and an excitation along B1-B2 and measurement across A1-A2 (see Fig. 3(b)). The measured $B_{eq}$ in pT/√Hz is fitted as follows,

$$B_{eq} = a_0 + a_1 \frac{1}{f^\gamma} \quad (6)$$

where, $a_0$, $a_1$ and $\gamma$ are the fit parameters. Usually, the exponent $\gamma$ value remains close to 1. The fit parameters and the extracted equivalent magnetic noise at 0.1 Hz, 1 Hz and 10 Hz are exhibited in Table I. Equivalent magnetic noise levels as low as 280 and 900 pT/√Hz at 1 Hz and 0.1 Hz respectively are obtained for this sensor.

Similarly, for the three crossing ellipses sensor, suitable ac excitation can be applied between any one of the three pairs of leads, C1-C2, D1-D2 or E1-E2. Here also, the field component perpendicular to the remanent magnetization lying along the major axes can be measured by measuring the PHE signal perpendicular to the magnetization. For instance, if the excitation is between D1 and D2, measuring the PHE signal either across the C1-E1 or E2-C2 will give the field component perpendicular to D1-D2. Likewise, field components perpendicular to C1-C2 and E1-E2 can be calculated by measuring the PHE signal between the leads perpendicular to the excitation line.

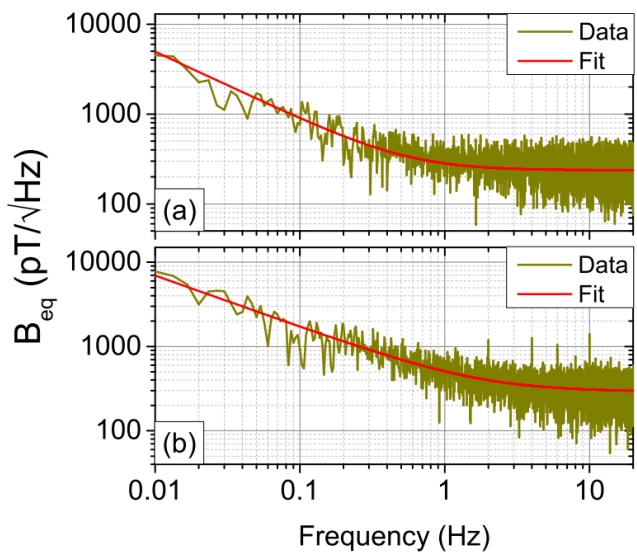

Fig. 3. Equivalent magnetic noise (Beq) vs frequency plot of 2-crossing ellipses sensor when the sensor is (a) excited along A1-A2 and the PHE signal is measured between B1-B2 and (b) excited along B1-B2 and measured between A1-A2 (refer Fig. 1(a) for lead captions).

The equivalent magnetic noise spectrum of 3-crossing ellipses sensor is shown in Fig.4 for excitation along (a) D1-D2, (b) E1-E2 and (c) C1-C2. Equivalent magnetic noise measured in pT/√Hz is fitted to Eq. (7). The EMN at different frequencies and the fit parameters are presented in Table I. EMN as low as 150 pT/√Hz at 1 Hz is obtained for this sensor when measured between C1-E1. These values are comparable to any of the existing commercially available MR magnetometers in the industry [17]. The EMN at 0.1 Hz is below 850 pT/√Hz and surpass that of the EMN of many CMOS and MEMS sensors [18],[19]. The equivalent magnetic noise at 10 Hz is around 100 pT/√Hz. We note variation in equivalent magnetic noise along different axes which is partially attributed to the fabrication and lithography and we believe that in the future with a higher quality deposition and fabrication processes, will get consistent results for all the axes. Both sensors (two and three ellipses) measure two components of the magnetic field in the overlap area of the ellipses with EMN level comparable to a single axis sensor [11] without the need to increase the sensor size.

We note that there are intrinsic advantages to increasing the number of crossing ellipses. As observed before, the higher the number of crossing ellipses the lower is the effective anisotropy field for each axis [12]. Moreover, the effective sensing volume increases due to increase in the overlap area. Both the decreased anisotropy field and increased sensing volume contribute to reduce the EMN (Eq. 4 and Eq. 5). Furthermore, the existence of multiple easy axes makes it possible to choose for each measurement the best axis (namely the one which is closest to being perpendicular to the direction of the signal) and measure with only one axis without loosing the large portion of the signal. Also, in the case of going from two to three crossing ellipses, we see that the contact resistance between the leads and the magnetic ellipses reduces significantly which increases the sensitivity. We note, however, that there is a limit to the gain from adding crossing ellipses as with increasing the number of ellipses, the sensor becomes magnetically less stable.

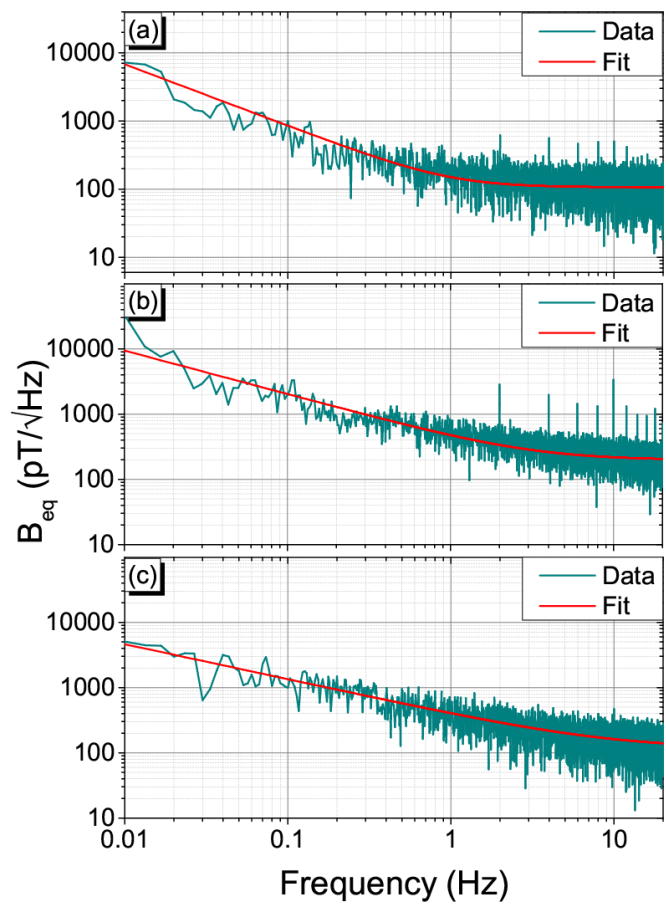

Fig. 4. Equivalent magnetic noise vs frequency plot of 3-crossing ellipses sensor when the sensor is excited along (a) D1-D2 and the PHE signal is measured between C1-E1, (b) excited along E1-E2 and measured between D1-C2 and (c) excited along C2-C1 and measured between E1-D2 (refer Fig. 1 for lead captions).

## IV. CONCLUSIONS

We demonstrate the performance of novel PHE sensors with two-, and three crossing ellipses as 2-axis magnetometers. These sensors can measure two axes of the magnetic field in the sensor plane while keeping the size and equivalent magnetic noise of single ellipse sensors. The magnetic field perpendicular to the excitation EA direction can be measured with equivalent magnetic noise better than 350 pT/√Hz at 1 Hz and around 900 pT/√Hz at 0.1 Hz. The obtained equivalent magnetic noises are attractive compared to the best commercial room-temperature MR magnetometers (see Supplementary Material). Note that, the different components of the magnetic field in present study are measured in exactly the same place, which is not exactly the case for the commercial 2-axis sensors. The simple design, low fabrication and integration cost, high performance and the smaller size make these novel PHE sensor useful for future device applications.

TABLE I
$B_{eq}$ AT DIFFERENT FREQUENCIES AND THE FIT PARAMETERS OF EQUATION (7) FOR TWO AND THREE CROSSING ELLIPSES SENSORS.

| Excitation | Measurement | Beq @0.1 Hz (pT/√Hz) | Beq @1 Hz (pT/√Hz) | Beq @10 Hz (pT/√Hz) | $a_0$ | $a_1$ | γ |
|---|---|---|---|---|---|---|---|
| Two Crossing ellipse sensors | | | | | | | |
| A1-A2 | B1-B2 | 900 | 280 | 246 | 154 | 237 | 0.75 |
| B1-B2 | A1-A2 | 1621 | 487 | 305 | 390 | 291 | 0.61 |
| Three Crossing ellipse sensors | | | | | | | |
| C1-C2 | D1-E2 | 1323 | 398 | 180 | 325 | 183 | 0.64 |
| C1-C2 | E1-D2 | 2035 | 478 | 218 | 423 | 310 | 0.70 |
| D1-D2 | C1-E1 | 826 | 150 | 103 | 106 | 107 | 0.90 |
| D1-D2 | E2-C2 | 785 | 185 | 123 | 140 | 123 | 0.72 |
| E1-E2 | D1-C2 | 1924 | 455 | 219 | 354 | 200 | 0.66 |
| E1-E2 | C1-D2 | 2119 | 523 | 320 | 423 | 310 | 0.69 |

## Acknowledgment

This work was supported in part by the (a) European Commission's HORIZON-TMA-MSCA-PF-EF under the Marie Skłodowska-Curie grant agreement No. 101106524, and (b) Planning and Budgeting Committee of the Council for Higher Education of Israel under Award vat/bat/cyc6/102.